\documentclass[aps,pre,preprint]{revtex4-1}
\usepackage{multirow,subeqnarray}
\usepackage[table]{xcolor}
\usepackage{rotating}
\usepackage{amsmath}
\usepackage{bm}
\usepackage{array}
\usepackage{graphicx}
\usepackage{dcolumn}   
\usepackage{amssymb}
\usepackage[english]{babel}
\def\r{{\bf r}}
\begin{document}
\title{Rotation of slender swimmers  in isotropic-drag media}
\date{\today}
\author{Lyndon Koens}
\email{lmk42@cam.ac.uk}
\author{Eric Lauga}
\email{e.lauga@damtp.cam.ac.uk}
\affiliation{ Department of Applied Mathematics and Theoretical Physics, University of Cambridge, Wilberforce Road, Cambridge CB3 0WA, United Kingdom}
\begin{abstract} 
The drag anisotropy of slender filaments is a critical physical property allowing  swimming in low-Reynolds number flows, and without it linear translation is impossible. Here we show that, in contrast, net rotation can occur under isotropic drag. We first demonstrate this result formally by considering the consequences of the force-  and torque-free conditions on swimming bodies and we then illustrate it with two examples (a simple swimmers made of three rods and a model bacterium with two helical flagellar filaments). Our results highlight the different role of hydrodynamic forces in generating translational vs.~rotational propulsion.
\end{abstract}
\maketitle
\def\v{\vspace{2cm}}

Since the 1950's \cite{Taylor1951,GRAY1955}, a continuous dialogue between theory and experiments has allowed us to unravel the fundamental physics of microorganism locomotion  \cite{Lauga2009} and we can now predict how different organisms swim \cite{Chattopadhyay2006,SMITH2009}, how they move to favorable environments \cite{Locsei2007,Drescher2010} and how they respond to boundaries \cite{Spagnolie2012,Denissenko2012a}. 
 The scientific community has also created its own series of synthetic microswimmers and attempted to optimise them  \cite{Zhang2010,Keaveny2013,Koens2016,Maggi2015a}.

Our understanding is made possible by our combined ability to (a) accurately measure the motion of micro-scale swimmers  \cite{Chattopadhyay2006,Drescher2010,Denissenko2012a} and (b)  theoretically describe the motion of the surrounding fluid through the incompressible Stokes equations \cite{Lauga2009, Leal2007}. The flow around the swimmer is obtained by enforcing that the fluid velocity on its surface is the same as the  velocity of the swimmer itself (no-slip boundary condition) and  the swimming kinematics are such that there is no net force and torque on the swimmer (free-swimming conditions).  Since the Stokes equations are linear and time independent, net propulsion can only be created by a stroke kinematics which breaks the time symmetry of the system, termed non-reciprocal  \cite{Purcell}.  In the vast majority of cases, microorganisms generate non-reciprocal strokes by sending bending waves \cite{Gaffney2011} or rotating \cite{Lauga2016} slender filaments termed flagella.

Such slender filaments are able to generate net propulsive forces due to their drag anisotropy at low Reynolds numbers (so-called drag-based thrust). Specifically, the drag per unit length acting on the slender filament is smaller for a translation along its centreline than for translation perpendicular to it. This is a fundamental property of small-scale fluid mechanics, which originates from the Green's function, $\bf G (\r)$, for the incompressible Stokes equations 
due to a point force $\mathbf{f}$ located at $\r_0$  \cite{Chwang2006}
\begin{equation}
\mathbf{G} (\r)= \frac{1}{8\pi\mu}\frac{\mathbf{I} + \mathbf{\hat{R}\mathbf{\hat{R}}}}{|\mathbf{R}|} \cdot \mathbf{f},
\end{equation}
where $\mu$ is  the  dynamic viscosity of the fluid, $\mathbf{I}$ is the identity tensor and $\mathbf{R}=\r-\r_0$ is the vector pointing from the location of the point force to the point of interest ($\bf\hat{R}$ is a unit vector in the same direction). Clearly, the flow resulting from the Green's function at any point in which $\mathbf{R}$ is parallel to $\mathbf{f}$ is twice as strong as the flow at a point with $\mathbf{R}$ perpendicular to $\mathbf{f}$ for the same $|\mathbf{R}|$.

The resulting modeling approaches to describe the motion of slender filaments in viscous fluids therefore also display this feature of drag-anisotropy. The  two commonly-used theories are resistive-force (or local-drag) theory, which is analytical but only logarithmically correct \cite{GRAY1955,Cox,Batchelor2006},  and slender-body theory, which has to be implemented numerically in general but is algebraically correct \cite{1976,Johnson1979,Koens2016}. 

In a number of important situations, the anisotropy of the drag is less prominent. For example some eukaryotic microorganisms have evolved hairs along their flagella called mastigonemes \cite{Brennen1975, Tottori2013}. These change the drag characteristics of the filament by making the drag parallel to the filament similar to, or smaller than, the drag perpendicular to the filament, allowing these swimmers to swim `backwards'. When these hairs are at the correct length and density the drag on the filament therefore becomes isotropic, removing the anisotropic influence imparted by the Stokes Green's function. Similarly in non-Newtonian environments the drag changes in complex ways \cite{Phan-Thien2012}. For example the drag on rods in a shear thinning fluid was seen to decrease as the non-Newtonian nature of the fluid was increased \cite{Chhabra2001}. This relative decrease was seen to be similar for motions parallel and perpendicular, though if the decrease was different for the two motions the system may again develop isotropic drag characteristics. Similarly if the Stokes Green's function was made isotropic (i.e. proportional to the Laplacian Green's function), the drag on a filament would arise from a line distribution of these isotropic Green's functions and would be isotropic as a result.
 
 \begin{figure}
\centering
\includegraphics[width=0.5\textwidth]{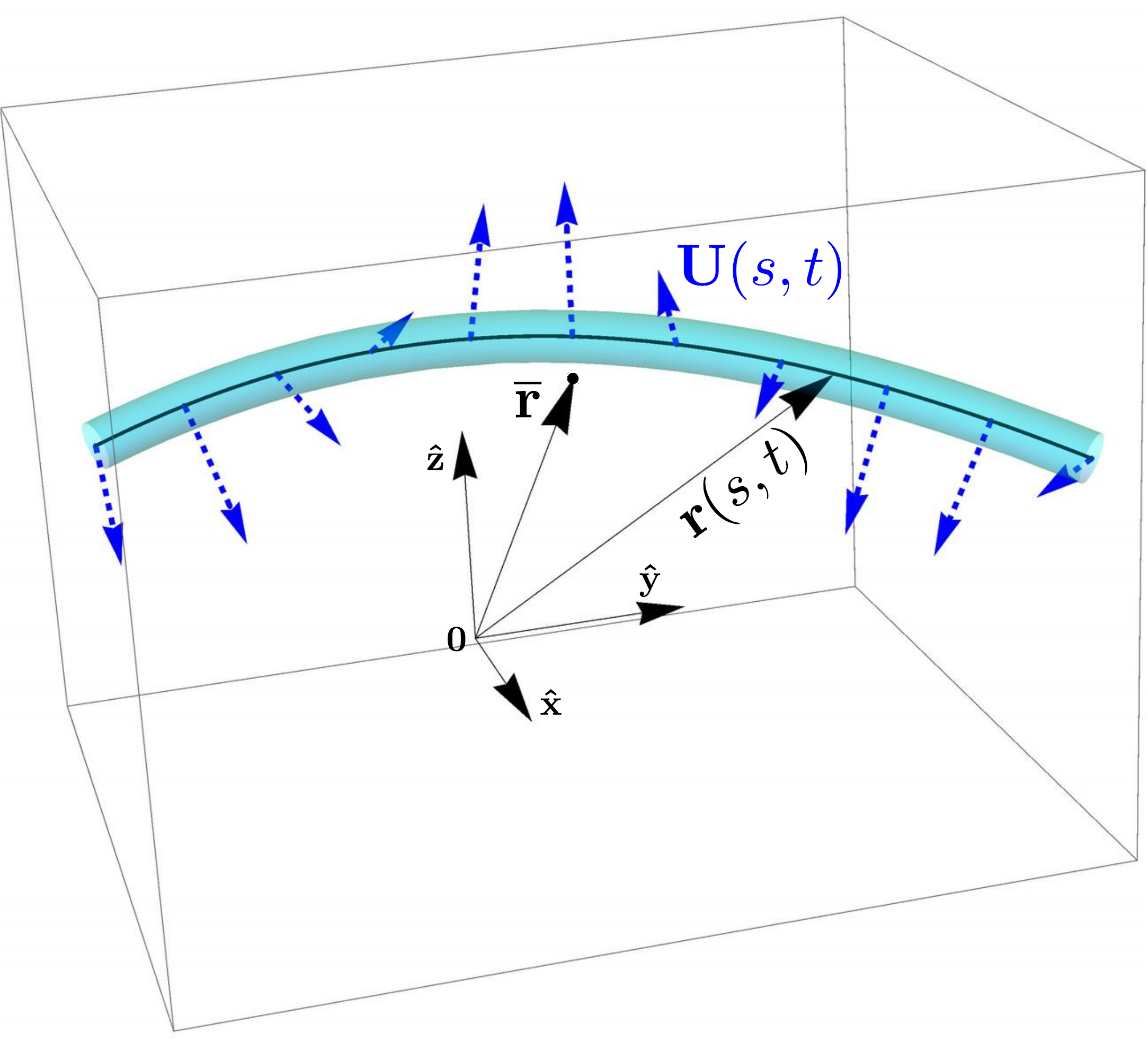}
\caption[Diagram showing the slender filament used to prove the necessity of drag anisotropy]{Diagram showing the slender filament used to prove the necessity of drag anisotropy. The black line represents the filaments centreline, $\mathbf{r}(s,t)$, and the blue dotted arrows indicate the instantaneous velocity of the filament at arclength $s$ and time $t$, $\mathbf{U}(s,t)$. The mean geometric position, $\mathbf{\overline{r}}(t)$, is also indicated for said filament.}
\label{fig:filamentimage}
\end{figure}

However, drag anisotropy is the critical physical ingredient to allow motion of the geometric center of a swimmer, and without it a net translation is impossible \cite{BECKER2003}. Indeed, consider an inextensible \cite{Pak2011} filament of length $L$ described by the centerline location $\mathbf{r}(s,t)$ and deforming its shape with the instantaneous velocity $\mathbf{U}(s,t)$ in the laboratory frame of reference (Fig.~\ref{fig:filamentimage}).   
Under isotropic drag, the hydrodynamic force density along the filament, at arclength $s$, scales as $\mathbf{f}(s,t) \propto{\bf U}(s,t)$, so that the velocity of the mean filament position, $\overline{\mathbf{r}}(t) = (\int \mathbf{r}(s,t) \,ds) /L$, is given by
\begin{eqnarray} 
\notag \frac{d \overline{\mathbf{r}}}{d t}& =& \frac{1}{L} \frac{d }{d t} \int_{0}^{L} \mathbf{r}(s,t) \,ds 
= \frac{1}{L}\int_{0}^{L} \mathbf{U}(s,t) \,ds\\ 
&&\propto \int_{0}^{L} \mathbf{f}(s,t) \,ds =\mathbf{0},
\label{rigidlinear}
\end{eqnarray}
since the swimmer is force-free at all times. Allowing the body to be extensible can break this condition, prompting the creation of many popular theoretical models, like extensible filament swimming \cite{Pak2011} and three sphere swimmers \cite{Najafi2004,Curtis2013}. However for many microswimmers, which are inextensible, drag anisotropy is a fundamental constraint on whether an organism can translate at low Reynolds number. 

The argument shown in Eq.~\eqref{rigidlinear} applies to the swimmer's translation. It is unclear if a similar reasoning may be used to rule out a net rotation. In this paper we show that, in fact, drag anisotropy is not required to generate rotation. This is first shown formally by considering the force- and torque-free condition for the  arbitrary deformation of a swimmer actuating slender appendages. We then illustrate the generation of rotation using model two swimmers: a lopsided paddle swimmer composed of three rods and a model bacterium with two flagellar filaments. Our work demonstrates that geometry alone can generate the conditions required to induce rotation and highlights the different role of hydrodynamic forces in generating translational vs.~rotational propulsion.
 
Using the notation above to describe the filament, we employ $\mathbf{U}_{\rm def}(s,t)$ to denote the instantaneous zero-mean deformation of the body in the laboratory frame.  We pick the instantaneous origin of the frame of reference to be the center of mass of swimmer and thus write 
 $\overline{\mathbf{r}}=\mathbf{0}$. 
The  velocity along the shape of the swimmer is therefore written as
\begin{equation}
\mathbf{U}(s,t) = \mathbf{U}_{\rm def}(s,t) + \mathbf{U}(t)+ \boldsymbol{\Omega}(t) \times \mathbf{r}(s,t),
\end{equation}
where $\mathbf{U}(t)$ and $\boldsymbol{\Omega}(t)$ are the instantaneous translation and rotational velocities.  If  motion occurs  is in a medium with isotropic drag, then Eq.~\eqref{rigidlinear} shows that $\mathbf{U}(t)=\mathbf{0}$ for all times.  Denoting  $\mathbf{f}(s,t) = \zeta \mathbf{U}(s,t)$ the isotropic relationship between the hydrodynamic  force density acting on the fluid and the velocity of the swimmer centerline, we  can write
\begin{equation}
\mathbf{f}(s,t) =  \zeta \mathbf{U}_{\rm def}(s,t)+ \zeta \boldsymbol{\Omega} (t) \times \mathbf{r}(s,t),
\end{equation}
where $\zeta$ is the isotropic drag per unit length. The form of $\zeta$ depends on the specific situation, whether that be  a non-Newtonian fluid or a complex geometry and it is left general here to demonstrate this affect in any environment where isotropic drag on a `filament' is present. From the isotropic drag condition, the torque density, $ \boldsymbol{\ell}(s,t)$, is given by
\begin{equation}\label{5}
 \boldsymbol{\ell}(s,t) =  \mathbf{r}(s,t)\times \mathbf{f}(s,t) +\gamma \mathbf{\hat{t}} \mathbf{\hat{t}} \cdot\boldsymbol{\Omega}(t)  ,
\end{equation} 
where $\mathbf{\hat{t}}\equiv {\bf \hat t}(s,t)$ is the tangent vector along the body centreline. The last term in Eq.~\eqref{5} accounts for the torque generated from local rotation about the centerline of the filament, with a rotational drag coefficient denoted  $\gamma$ (in both  illustrative examples below we chose $\gamma=0$, but it has been left here for completeness).   We thus get  a torque density given by
\begin{eqnarray}
\notag \boldsymbol{\ell}(s,t) &= &     \zeta \mathbf{r}(s,t)\times \left[\mathbf{U}_{\rm def}(s,t)+\boldsymbol{\Omega}(t) \times \mathbf{r}(s,t)  \right] \\
 &&+ \gamma \mathbf{\hat{t}} \mathbf{\hat{t}} \cdot\boldsymbol{\Omega}(t).
\end{eqnarray} 
The total force and torque on the body are then given by
\begin{eqnarray}
\label{7}
\mathbf{0} &=\displaystyle \int_{0}^{L}\mathbf{f}(s,t) \,ds   =& \zeta \boldsymbol{\Omega}(t) \times \int_{0}^{L} \mathbf{r}(s,t) \,ds , \\
 \mathbf{0}& =\displaystyle\int_{0}^{L} \boldsymbol{\ell}(s,t) \,ds =&  \zeta \int_{0}^{L}\mathbf{r}(s,t)\times\mathbf{U}_{\rm def}(s,t)\,ds \label{8}\\
&& + \int_{0}^{L} \left( \zeta \mathbf{I} r^{2} - \zeta \mathbf{r}\mathbf{r} + \gamma \mathbf{\hat{t}} \mathbf{\hat{t}}\right) \,ds \cdot \boldsymbol{\Omega}(t),
\notag\end{eqnarray} 
where we have used the vector  identity $\mathbf{a}\times(\mathbf{b}\times\mathbf{c}) =  \mathbf{b} (\mathbf{a}.\mathbf{c}) -\mathbf{c} (\mathbf{a}.\mathbf{b})$. 
The zeros on the left hand side of Eqs.~\eqref{7}-\eqref{8} reflect the fact that the swimmer is force-  and torque-free for all times. The force free condition is automatically satisfied since $\bf \bar  r = 0$.   The torque-free condition, Eq.~\eqref{8}, leads to an explicit equation for the rotation rate of the swimmer as
\begin{eqnarray}\label{torque}
 \mathbf{R}(t)\cdot \boldsymbol{\Omega}(t) &=& - \zeta \int_{0}^{L}\mathbf{r}(s,t)\times\mathbf{U}_{\rm def}(s,t)\,ds,  \label{genrot}
\end{eqnarray}
where the resistance tensor, ${\bf R}(t)$, is instantaneously given by
\begin{equation} \label{resistance}
\mathbf{R}(t) = \int_{0}^{L} \left( \zeta \mathbf{I} r^{2} - \zeta \mathbf{r}\mathbf{r} + \gamma \mathbf{\hat{t}} \mathbf{\hat{t}}\right) \,ds.
\end{equation}
Hence, provided the deformation generates a net torque on the right-hand side of Eq.~\eqref{torque}, and since
the resistance tensor $\bf R$ is always positive definite for any finite-length filament, 
then $\boldsymbol{\Omega} (t)\neq \mathbf{0}$ without the need for drag anisotropy. Note that this derivation is independent of the reference frame as it involves instantaneous velocities (the laboratory frame can be chosen to match the swimming frame at the time $t$, without loss of generality).

  Physically, anisotropy is necessary to create any motion. In the case of linear velocity this anisotropy must come from the drag, as demonstrated by Eq.~\eqref{rigidlinear}. However in the case of rotation, the configuration of the centerline, $\mathbf{r}(s,t)$, can generate the required anisotropy, as shown by the non-isotropic resistance tensor ${\bf R}_{c}$, even in the presence of isotropic drag.  This instantaneous rotation can then generate a net rotation over a period if the deformation undergoes a non-reciprocal stroke, as per the scallop theorem \cite{Purcell}.

\begin{figure}
\begin{center}
\includegraphics[width=.5\textwidth]{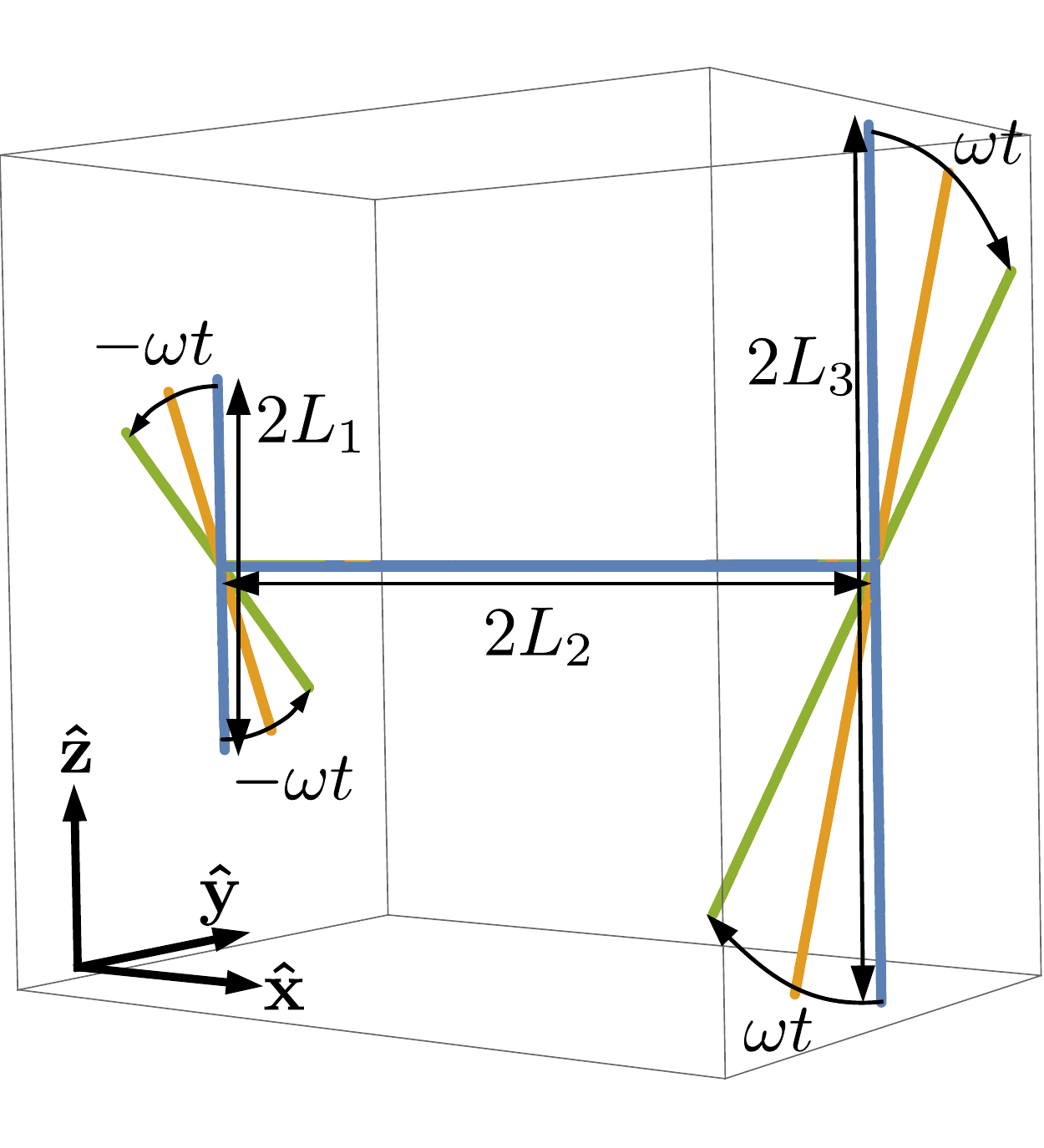}
\caption{(Color online) Representation of the lopsided paddle swimmer with three arms of lengths $2L_1$, $2L_2$ and $2L_3$. The different colored arms show the swimmer configuration at different times with  arrows indicating their rotation. The frame of reference $(x,y,z)$ rotate with the swimmer so that the $x$ axis is instantaneously aligned with rod \#2.}
\label{fig:paddleswimmer}
\end{center}
\end{figure}

To provide further intuition, we illustrate this result on two simple examples exhibiting non-zero angular velocities. Note that many more examples can be created from existing calculations that use resistive-force theory by formally setting the drag to be isotropic \cite{Chattopadhyay2006, Friedrich2010}.

 First we consider an elementary lopsided paddle swimmer as  shown in Fig.~\ref{fig:paddleswimmer}. This swimmer consists of three straight rods of lengths $2 L_{1}$, $2 L_{2}$ and $2 L_{3}$. The first and third rods are perpendicular to the second rod and both positioned at opposite ends of the second. The first and third rod then rotate around the  axis of the second rod in opposing directions with period $T=2\pi/\omega$. The centerline of this swimmer is described for all times and in the $(x,y,z)$ frame rotating with it (see Fig.~\ref{fig:paddleswimmer}) as
\begin{equation}
\mathbf{r} + \overline{\mathbf{r}} = \left\{ \begin{array}{c c}
\left\{ - L_{2} , - s_{1}\sin \omega t, s_{1}\cos \omega t \right\} & - L_{1} < s_{1} < L_{1}, \\
\left\{s_{2},0,0 \right\} & - L_{2} < s_{2} < L_{2}, \\
\left\{ L_{2} , s_{3}\sin \omega t, s_{3}\cos \omega t \right\} & - L_{3} < s_{3} < L_{3},
\end{array} \right.
\end{equation}
where $s_{i}$ describes the configuration of rod $i$.
The origin of the reference frame is located at the center of the swimmer, $\overline{\mathbf{r}}$, which is found by
\begin{eqnarray}
\notag 2 (L_{1} + L_{2}+ L_{3}) \overline{\mathbf{r}} &=& \int (\mathbf{r} + \overline{\mathbf{r}}) \,ds
\\
\notag &=&  \int_{-L_{1}}^{L_{1}} \left\{ - L_{2} , - s_{1}\sin \omega t , s_{1}\cos \omega t \right\} \,ds_{1} \\ 
\notag &&+ \int_{-L_{2}}^{L_{2}} \left\{s_{2},0,0 \right\}\,ds_{2}  \\
\notag && +\int_{-L_{3}}^{L_{3}} \left\{ L_{2} , s_{3}\sin \omega t , s_{3}\cos \omega t  \right\} \,ds_{3} \\
&=& \left\{2 L_{2}(L_{3}-L_{1}),0,0\right\}.
\label{center}\end{eqnarray}
Hence the center of the  swimmer is a point on the second rod closer to the longest rotating rod.
 The deformation velocity  for this swimmer is given by
\begin{equation}
\mathbf{U}_{\rm def} = \omega \left\{ \begin{array}{c c}
\left\{ 0, - s_{1}\cos \omega t , -s_{1}\sin \omega t \right\} & - L_{1} < s_{1} < L_{1}, \\
\{0,0,0\} & - L_{2} < s_{2} < L_{2}, \\
\left\{ 0 , s_{3}\cos \omega t , -s_{3}\sin \omega t \right\} & - L_{3} < s_{3} < L_{3},
\end{array} \right. 
\end{equation}
and it generates an instantaneous net force and torque on the fluid of magnitudes
\begin{eqnarray}
\mathbf{F} &=&  \zeta \int \mathbf{U}_{\rm def} \,ds = \{0,0,0\}, \\
\mathbf{L} &=&  \zeta \int \mathbf{r}\times\mathbf{U}_{\rm def} \,ds =\left\{ -\frac{2}{3} (L_{3}^{3}-L_{1}^{3}) \zeta \omega,0,0\right\},
\end{eqnarray}
where the integrals are taken over all three rods as in Eq.~\eqref{center}.  Assuming for simplicity that $\gamma=0$, then the resistance matrix from Eq.~\eqref{resistance} is instantaneously 
\begin{equation}
\mathbf{R} (t) = \zeta\left(\begin{array}{c c c}
\frac{2}{3} (L_{1}^{3}+L_{3}^{3}) &0 &0 \\
0&  \frac{2}{3}A(t) & B(t) \\
0& B(t)& \frac{2}{3}C(t) \\
\end{array} \right),
\end{equation}
where the coefficients $A$, $B$, and $C$ are given by
\begin{eqnarray}
\notag A(t) &=& \frac{L_{2}^{2} [4 L_{1}(L_{2}+3 L_{3})+L_{2}(L_{2}+4L_{3})]}{L_{1}+L_{2}+L_{3}} \\
&& +(L_{1}^{3}+L_{3}^{3}) \cos^{2} \omega t,  \\
B(t) &=& -\frac{1}{3}(L_{3}^{3}-L_{1}^{3})\sin 2 \omega t,\\
\notag C(t) &=& \frac{L_{2}^{2} [4 L_{1}(L_{2}+3 L_{3})+L_{2}(L_{2}+4L_{3})]}{L_{1}+L_{2}+L_{3}} \\
&& +(L_{1}^{3}+L_{3}^{3}) \sin^{2}  \omega t.
\end{eqnarray}
Inverting $\bf R$, we obtain that the instantaneous rotation rate of the lopsided paddle swimmer is constant and given by
 \begin{equation}
 \boldsymbol{\Omega} = \mathbf{R}(t)^{-1}\mathbf{L} = \left\{- \frac{L_{3}^{3}-L_{1}^{3}}{L_{3}^{3}+L_{1}^{3}} \omega,0,0\right\},
 \end{equation}
which has a nonzero angular velocity provided $L_{1}$ is not equal to $L_{3}$. Clearly, if both rods have finite sizes, then the rotation rate of the whole swimmer is not equal to minus the rotation rates of each rod and indicates that a net rotation of the whole swimmer body can be induced purely from geometry.

\begin{figure}
\begin{center}
\includegraphics[width=.6\textwidth]{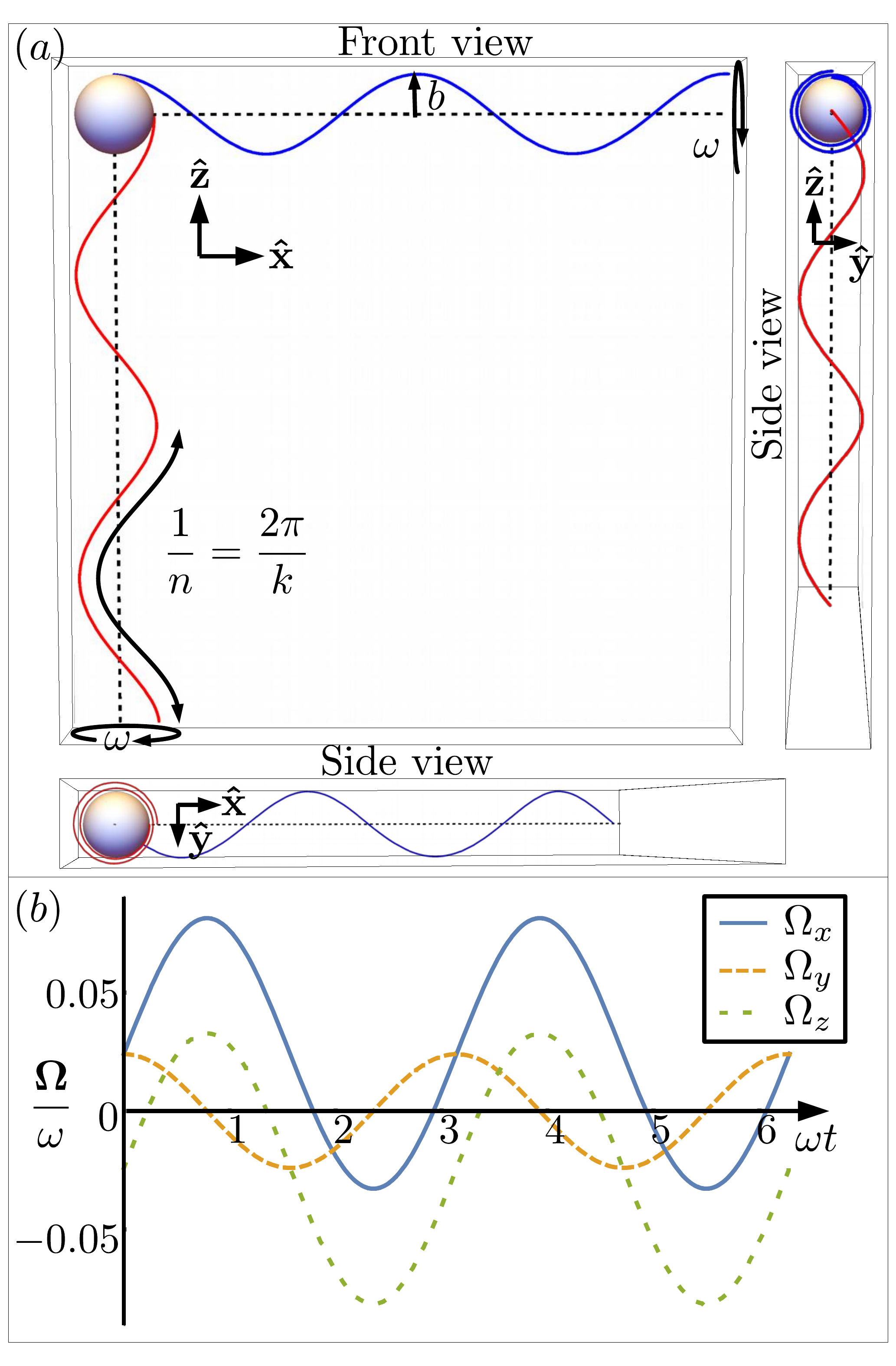}
\caption{Rotation of a model bacterium under isotropic drag. (a): Geometry of the  swimmer composed of two identical helices of perpendicular axis,  amplitude $b$ and wavenumber $k$ and rotate with frequency $\omega$. The  cell body is small compared to the helices and can thus be ignored. (b): Components of the angular velocity of the swimmer in the swimmer frame, $\boldsymbol{\Omega}$ (scaled by $\omega$),  as a function of dimensionless time. 
For all figures $b=0.05$ and $k=4 \pi$.}
\label{fig:helix}
\end{center}
\end{figure}

As a second example we consider a model for a bacterium with two flagellar filaments, as illustrated in Fig.~\ref{fig:helix}a. The swimmer is composed of two identical rigid helices  attached at one end to a cell body and  oriented with their helix axes perpendicular to each other. Each  helix rotates around its axis with period $T=2\pi/\omega$. For simplicity we ignore hydrodynamically the  presence of the cell body which is  correct in the limit where the helical flagella are much longer than the body. In the frame of reference attached to the swimmer (see Fig.~\ref{fig:helix}a), the location of each helix is  given by
\begin{eqnarray}
\mathbf{r}_{1} + \overline{\mathbf{r}} &=& \left\{\alpha s, b  \sin(k x -\omega t), b  \cos(k x - \omega t) \right\},  \\
\mathbf{r}_{2} + \overline{\mathbf{r}} &=& \left\{b  \cos(k x - \omega t), b  \sin(k x -\omega t), -\alpha s  \right\},
\end{eqnarray}
where $\mathbf{r}_{1}$ and $\mathbf{r}_{2}$ denote, respectively, the centerlines of the   first and second helices,  $0<s<1$, $k=2 n \pi$  is the wavenumber of the helix for a positive integer $n$, $b$ its amplitude, and $\alpha$ is the cosine of the helix angle which satisfies  $\alpha^{2} + b^{2} k^{2} =1$ for  inextensible helices.   This swimmer rotates about the centre $\overline{ \bf r}$ defined as
\begin{eqnarray}
\notag  \overline{\mathbf{r}} &=& \frac{1}{2}\int_{0}^{1} (\mathbf{r}_{1} +\overline{\mathbf{r}}) \,ds + \frac{1}{2}\int_{0}^{1} (\mathbf{r}_{2} +\overline{\mathbf{r}}) \,ds \\
&=& \frac{\alpha}{4}\{1,0,-1\},
\end{eqnarray}
which is  the origin for both $\mathbf{r}_{1}$ and $\mathbf{r}_{2}$. The deformation velocity for 
 for each helix,  $\mathbf{U}_{{\rm def},i}$,  is then obtained  by  computing  $\partial_{t}\mathbf{r}_{i}$ ($i=1,2$). We obtain that the  rotation of the helices around their axes generates net forces and torques on the body of magnitude
\begin{eqnarray}
\mathbf{F}(t) &=&  \zeta \int \mathbf{U}_{\rm def} \,ds = \{0,0,0\}, \\
\mathbf{L}(t) &=&  \zeta \int \mathbf{r}\times\mathbf{U}_{\rm def} \,ds  \\
&=& \frac{b \zeta \omega}{2 n \pi} \left\{ 2 b n \pi +\alpha \sin \omega t,\alpha \cos \omega t , \alpha \sin \omega t-2 b n \pi \right\}. \notag 
\end{eqnarray}
Assuming $\gamma=0$ for simplicity, the resistance matrix relating torque and rotation for this bacterial configuration can be computed exactly and we obtain
\def\dd{\displaystyle}
\begin{equation}
{\mathbf{R}}(t) = \zeta\left(\begin{array}{c c c}
\dd \frac{36 b^{2} +5 \alpha^{2}}{24}  &\dd\frac{b \alpha \cos \omega t }{2 n \pi} &\dd-\frac{\alpha^{2}}{8} \\
\dd\frac{b \alpha \cos \omega t }{2 n \pi} &  b^{2} +\dd \frac{5}{12} \dd\alpha^{2} & \dd-\frac{b \alpha \cos \omega t }{2 n \pi} \\
\dd-\frac{\alpha^{2}}{8} &\dd -\frac{b \alpha \cos \omega t }{2 n \pi}& \dd\frac{36 b^{2} +5 \alpha^{2}}{24}  \\
\end{array} \right),
\end{equation}
and the torque-free condition then gives the swimmer an instantaneous angular velocity of
\begin{equation}
\boldsymbol{\Omega} (t)=- 6 b \omega \left\{  A_{2}(t) + B_{2}(t), C_{2}(t), A_{2}(t) -B_{2}(t) \right\},
\end{equation}
where
\begin{eqnarray}
A_{2} (t)&=& \frac{\alpha \sin \omega t}{n \pi (18 b^{2} + \alpha^{2})}, \\
B_{2} (t)&=& \frac{2 b n^{2} \pi^{2} (12 b^{2}+5 \alpha^{2}) - 6 b \alpha^{2} \cos^{2} \omega t }{n^{2} \pi^{2} ( 108 b^{4} + 69 b^{2} \alpha^{2} +10 \alpha^{4}) - 36 b^{2} \alpha^{2} \cos^{2} \omega t },  
\quad\quad \\
C_{2} (t)&=& \frac{  2 n \pi \alpha (3 b^{2} +2 \alpha^{2})\cos \omega t }{n^{2} \pi^{2} ( 108 b^{4} + 69 b^{2} \alpha^{2} +10 \alpha^{4}) - 36 b^{2} \alpha^{2} \cos^{2} \omega t }. 
\end{eqnarray}

\begin{figure}
\centering
\includegraphics[width=\textwidth]{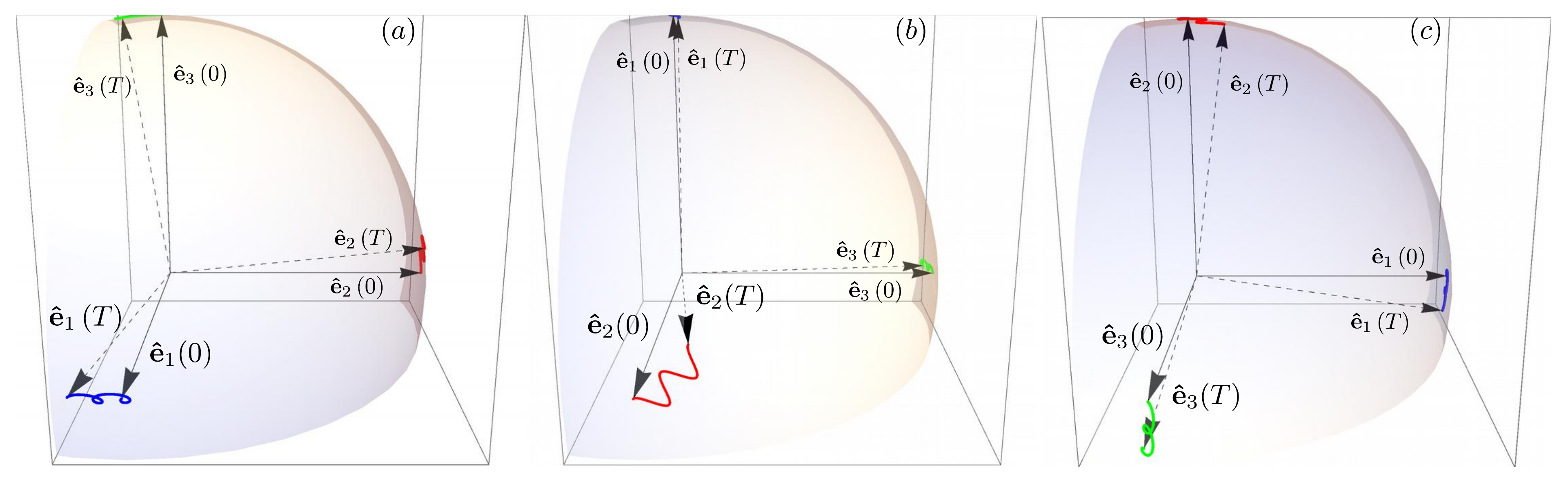}
\caption[Trajectories of the body vectors on the surface of a unit sphere]{Trajectories of the body vectors ($\mathbf{\hat{e}}_{1}(t)$, $\mathbf{\hat{e}}_{2}(t)$, $\mathbf{\hat{e}}_{3}(t)$) on the surface of a unit sphere ($r_{h}=0.05$, $k=4 \pi$). (a) The trajectory of $\mathbf{\hat{e}}_{1}(t)$ (blue); (b) The trajectory of $\mathbf{\hat{e}}_{2}(t)$ (red); (c) The trajectory of $\mathbf{\hat{e}}_{3}(t)$ (green). The paths over the sphere represent the motion with the initial (solid) and final (dashed) vectors clearly labelled. In the above $T= 2 \pi /\omega$.}
\label{fig:rotationswimmer}
\end{figure}

We plot in  Fig.~\ref{fig:helix}b all the components of  $\boldsymbol{\Omega}$, nondimensionalized by $\omega$, for the values $b=0.05$ and $k=4\pi$ (i.e.~$n=2$). We see that the  instantaneous rotational velocity is non-zero for all body directions and oscillates sinusoidally, around a mean value of 
\begin{equation}
 \frac{\omega}{2 \pi} \int_{0}^{{2\pi}/{\omega}}\boldsymbol{\Omega} \,dt = \omega \left[\left(1-\sqrt{W}\right) +\frac{6 r_{h}^{2}}{9 r_{h}^{2}+2 \alpha^{2} }\sqrt{W}\right] \{-1 ,0,1\},
\end{equation}
where 
\begin{equation}
W = \frac{108 r_{h}^{2} + \alpha^{2} \left(69-108 r_{h}^{2}\right)- \alpha^{4} \left(69 -10 n^{2} \pi^{2}\right)}{108 r_{h}^{2} + \alpha^{2} \left(69-72 r_{h}^{2}\right)- \alpha^{4} \left(69 -10 n^{2} \pi^{2}\right)}\cdot
\end{equation} 
The angular displacement experienced in the laboratory frame can be found from
\begin{equation}
\frac{d \mathbf{\hat{e}}_{i}}{d t} = \boldsymbol{\Omega}(t) \times \mathbf{\hat{e}}_{i},
\end{equation}
where $\mathbf{\hat{e}}_{i}$ is the body frame vector $i=1,2,3$. We take $\mathbf{\hat{e}}_{1}(t)$ to be the body vector aligned the helix axis of $\mathbf{r}_{1}$,  $\mathbf{\hat{e}}_{3}(t)$ the  body vector aligned with the helix axis of $\mathbf{r}_{2}$ and $\mathbf{\hat{e}}_{2}(t)$ the body vector perpendicular to the helix axes. The above equation was solved numerically for a bacterial swimmer with $r_{h} = 0.05$ and $k=4\pi$ \cite{Koens2014}. In this configuration, the body vectors, after one period of rotation, become
\begin{eqnarray}
\mathbf{\hat{e}}_{1}\left(T =\frac{2 \pi}{\omega} \right) &=& \{0.989.-0.149,-0.003 \}\\
\mathbf{\hat{e}}_{2}\left(T =\frac{2 \pi}{\omega} \right) &=& \{0.148, 0.9778,0.149 \} \\
\mathbf{\hat{e}}_{3}\left(T =\frac{2 \pi}{\omega} \right) &=& \{-0.020,-0.148, 0.989 \}
\end{eqnarray}
where we have assumed $\mathbf{\hat{e}}_{1}(0) = \{1,0,0\}$, $\mathbf{\hat{e}}_{2}(0) = \{0,1,0\}$, and $\mathbf{\hat{e}}_{3}(0) = \{0,0,1\}$. These vectors are written in terms of the $x$, $y$, and $z$ coordinates of the laboratory frame. Fig.~\ref{fig:rotationswimmer} plots the trajectories of these vectors on the surface of a unit sphere. Hence under isotropic drag, this model bacterium undergoes a non-trivial net rotation in the laboratory frame due solely to the anisotropy of its shape. 

Anisotropy in their linear drag is a  requirement for microorganisms using filamentous appendages to undergo net translation through  viscous fluids.
 By extension one may think that such drag anisotropy is also required to generate rotation. In this paper we showed that in fact rotation was possible in a system with isotropic drag. This result was evident when considering the force- and torque-free conditions directly and allowed us to demonstrate  that in such a fluid the rotation is physically generated by the anisotropy  in the  shape of the swimmer (specifically, the centerline of the slender filaments it actuates). We derived this rotation for an arbitrary body and then illustrated it on two model swimmers, a lopsided paddle swimmer and a multi-flagellated model bacterium, which both exhibit non-zero rotation within an isotropic-drag medium. Other examples may also be created by making the drag isotropic in existing resistive force theory studies \cite{Chattopadhyay2006, Friedrich2010}.  Our results highlight the different role of hydrodynamic forces in generating translational vs.~rotational propulsion  and may change our understanding of the physical requirements for rotational motion in complex environments.

\paragraph*{Acknowledgements}
This research was funded in part by the European Union through a Marie Curie CIG grant (EL) and by the Cambridge Trust (LK).

\bibliographystyle{ieeetr}
\bibliography{library}

\begin{thebibliography}{10}

\bibitem{Taylor1951}
G.~Taylor, ``{Analysis of the swimming of microscopic organisms},'' {\em Proc.
  R. Soc. A Math. Phys. Eng. Sci.}, vol.~209, pp.~447--461, 1951.

\bibitem{GRAY1955}
J.~Gray and G.~J. Hancock, ``{The Propulsion of sea-urchin spermatozoa},'' {\em
  J. Exp. Biol.}, vol.~32, pp.~802--814, 1955.

\bibitem{Lauga2009}
E.~Lauga and T.~R. Powers, ``{The hydrodynamics of swimming microorganisms},''
  {\em Reports Prog. Phys.}, vol.~72, p.~096601, 2009.

\bibitem{Chattopadhyay2006}
S.~Chattopadhyay, R.~Moldovan, C.~Yeung, and X.~Wu, ``{Swimming efficiency of
  bacterium \textit{Escherichia coli}.},'' {\em Proc. Natl. Acad. Sci. U. S.
  A.}, vol.~103, pp.~13712--7, 2006.

\bibitem{SMITH2009}
D.~J. Smith, E.~A. Gaffney, J.~R. Blake, and J.~C. Kirkman-Brown, ``{Human
  sperm accumulation near surfaces: a simulation study},'' {\em J. Fluid
  Mech.}, vol.~621, p.~289, 2009.

\bibitem{Locsei2007}
J.~Locsei, ``{Persistence of direction increases the drift velocity of run and
  tumble chemotaxis.},'' {\em J. Math. Biol.}, vol.~55, pp.~41--60, 2007.

\bibitem{Drescher2010}
K.~Drescher, R.~E. Goldstein, and I.~Tuval, ``{Fidelity of adaptive
  phototaxis.},'' {\em Proc. Natl. Acad. Sci. U. S. A.}, vol.~107,
  pp.~11171--6, 2010.

\bibitem{Spagnolie2012}
S.~Spagnolie and E.~Lauga, ``{Hydrodynamics of self-propulsion near a boundary:
  predictions and accuracy of far-field approximations},'' {\em J. Fluid
  Mech.}, vol.~700, pp.~105--147, 2012.

\bibitem{Denissenko2012a}
P.~Denissenko, V.~Kantsler, D.~J. Smith, and J.~Kirkman-Brown, ``{Human
  spermatozoa migration in microchannels reveals boundary-following
  navigation.},'' {\em Proc. Natl. Acad. Sci. U. S. A.}, vol.~109,
  pp.~8007--10, 2012.

\bibitem{Zhang2010}
L.~Zhang, K.~E. Peyer, and B.~J. Nelson, ``{Artificial bacterial flagella for
  micromanipulation.},'' {\em Lab Chip}, vol.~10, pp.~2203--15, 2010.

\bibitem{Keaveny2013}
E.~E. Keaveny, S.~W. Walker, and M.~J. Shelley, ``{Optimization of chiral
  structures for microscale propulsion.},'' {\em Nano Lett.}, vol.~13,
  pp.~531--7, 2013.

\bibitem{Koens2016}
L.~Koens and E.~Lauga, ``{Slender-ribbon theory},'' {\em Phys. Fluids},
  vol.~28, p.~013101, 2016.

\bibitem{Maggi2015a}
C.~Maggi, J.~Simmchen, F.~Saglimbeni, J.~Katuri, M.~Dipalo, F.~{De Angelis},
  S.~Sanchez, and R.~{Di Leonardo}, ``{Self-assembly of micromachining systems
  powered by janus micromotors.},'' {\em Small}, vol.~12, p.~446, 2015.

\bibitem{Leal2007}
L.~G. Leal, {\em {Advanced Transport Phenomena: Fluid Mechanics and Convective
  Transport Processes}}.
\newblock Cambridge University Press, 2007.

\bibitem{Purcell}
E.~M. Purcell, ``{Life at low Reynolds number},'' {\em Am. J. Phys}, vol.~45,
  p.~3, 1977.

\bibitem{Gaffney2011}
E.~Gaffney, H.~Gad{\^{e}}lha, D.~Smith, J.~Blake, and J.~Kirkman-Brown,
  ``{Mammalian sperm motility: observation and theory},'' {\em Annu. Rev. Fluid
  Mech.}, vol.~43, pp.~501--528, 2011.

\bibitem{Lauga2016}
E.~Lauga, ``{Bacterial hydrodynamics},'' {\em Annu. Rev. Fluid Mech.}, vol.~48,
  pp.~105--130, 2016.

\bibitem{Chwang2006}
A.~T. Chwang and T.~Y.-T. Wu, ``{Hydromechanics of low-Reynolds-number flow.
  Part 2. Singularity method for Stokes flows},'' {\em J. Fluid Mech.},
  vol.~67, p.~787, 1975.

\bibitem{Cox}
R.~G. Cox, ``{The motion of long slender bodies in a viscous fluid Part 1.
  General theory},'' {\em J. Fluid Mech.}, vol.~44, p.~791, 1970.

\bibitem{Batchelor2006}
G.~K. Batchelor, ``{Slender-body theory for particles of arbitrary
  cross-section in Stokes flow},'' {\em J. Fluid Mech.}, vol.~44, p.~419, 1970.

\bibitem{1976}
J.~Lighthill, ``{Flagellar hydrodynamics: The John von Neumann lecture,
  1975},'' {\em SIAM Rev.}, vol.~18, pp.~161--230, 1976.

\bibitem{Johnson1979}
R.~E. Johnson, ``{An improved slender-body theory for Stokes flow},'' {\em J.
  Fluid Mech.}, vol.~99, p.~411, 1979.

\bibitem{Brennen1975}
C.~Brennen, ``{Locomotion of flagellates with mastigonemes},'' {\em J.
  Mechanochemistry Cell Motil.}, vol.~3, p.~207, 1975.

\bibitem{Tottori2013}
S.~Tottori and B.~J. Nelson, ``{Artificial helical microswimmers with
  mastigoneme-inspired appendages},'' {\em Biomicrofluidics}, vol.~7,
  p.~061101, 2013.

\bibitem{Phan-Thien2012}
N.~Phan-Thien, {\em {Understanding Viscoelasticity: An Introduction to
  Rheology}}.
\newblock Springer Science {\&} Business Media, 2012.

\bibitem{Chhabra2001}
R.~Chhabra, K.~Rami, and P.~Uhlherr, ``{Drag on cylinders in shear thinning
  viscoelastic liquids},'' {\em Chem. Eng. Sci.}, vol.~56, pp.~2221--2227,
  2001.

\bibitem{BECKER2003}
L.~E. Becker, S.~A. Koehler, and H.~A. Stone, ``{On self-propulsion of
  micro-machines at low Reynolds number: Purcells three-link swimmer},'' {\em
  J. Fluid Mech.}, vol.~490, pp.~15--35, 2003.

\bibitem{Pak2011}
O.~S. Pak and E.~Lauga, ``{Extensibility enables locomotion under isotropic
  drag},'' {\em Phys. Fluids}, vol.~23, p.~081702, 2011.

\bibitem{Najafi2004}
A.~Najafi and R.~Golestanian, ``{Simple swimmer at low Reynolds number: three
  linked spheres.},'' {\em Phys. Rev. E. Stat. Nonlin. Soft Matter Phys.},
  vol.~69, p.~062901, 2004.

\bibitem{Curtis2013}
M.~P. Curtis and E.~A. Gaffney, ``{Three-sphere swimmer in a nonlinear
  viscoelastic medium},'' {\em Phys. Rev. E}, vol.~87, p.~043006, 2013.

\bibitem{Friedrich2010}
B.~M. Friedrich, I.~H. Riedel-Kruse, J.~Howard, and F.~J{\"{u}}licher,
  ``{High-precision tracking of sperm swimming fine structure provides strong
  test of resistive force theory.},'' {\em J. Exp. Biol.}, vol.~213,
  pp.~1226--34, 2010.

\bibitem{Koens2014}
L.~Koens and E.~Lauga, ``{The passive diffusion of \textit{Leptospira
  interrogans}},'' {\em Phys. Biol.}, vol.~11, pp.~1--15, 2014.

\end{thebibliography}
\end{document}